\documentclass[12pt]{article}
\usepackage{graphicx}

\begin{document}

\title{\textbf{WIDE-FIELD CORRECTOR FOR A GREGORY
 TELESCOPE}\thanks{To be published in \textit{Astronomy~Reports}. Original Russian
text \copyright \,2006 by Terebizh.}}

\author{V.~Yu.~Terebizh\thanks{98409 Nauchny, Crimea, Ukraine;
 \,E-mail:\, \textsf{terebizh@crao.crimea.ua}}\\
 \small{\textit{Sternberg Astronomical Institute, Moscow, Russia}}}

\date{\footnotesize{Received September~10, 2005}}

\maketitle

\begin{quote}
\small{\textbf{Abstract}~--- A form of prime focus corrector for the Gregory
system is proposed that provides the sub-arcsecond field of view up to
$3^\circ$ in diameter for the spectral range $0.35-0.90$ microns. The corrector
includes five lenses made of same glass (fused silica is preferable). The
distinctive feature of the corrector consists in dissimilar use of the central
and edge zones of a front lens disposed in the exit pupil of a two-mirror
system.

As an example, the f/1.9 telescope is considered with the 6.5-m aperture and
the total length 8.8~m. Its primary and secondary mirrors are pure ellipsoids
close to concave paraboloid and concave sphere, respectively. In the basic
configuration, all surfaces of the corrector are spherical. The diameter of a
star image $D_{80}$ varies from $0''.25$ on the optical axis up to $0''.50$ at
the edge of the $2^\circ.3$ field. Only slightly worse images shows spherical
corrector for the $2^\circ.4$ field of view. The fraction of vignetted rays
grows on 1.7\% from the center of field to its edges. Aspherization of some
lens surfaces allows to reach sub-arcsecond images in the field of $3^\circ.0$
in diameter.\\
 \copyright \, \textit{2006 MAIK ``Nauka/Interperiodica''.}

\medskip

}
\end{quote}

\newpage
\section*{Introduction}

Recent designs of lens correctors for the large telescopes provide the field of
view of sub-arcsecond image quality up to $3^\circ$ in a Cassegrain system
(Hodapp et al., 2003), in a three-mirror Mersenn--Schmidt system (Paul, 1935;
Willstrop, 1984; Angel et al., 2000; Seppala, 2002), and in a prime focus of
hyperbolic mirror (Terebizh, 2004). In the two former cases, a telescope is
quite compact, while the surfaces of mirrors and lenses are complicated. In the
latter case, it is possible to achieve the above field even at all spherical
lenses, but the telescope length is about focal length of its primary mirror.

The Gregory system (both classic, with a paraboloidal primary, and aplanatic,
with ellipsoidal surfaces of the mirrors) has an attractive feature: its exit
pupil is not imaginary, as it takes place for the Cassegrain system, but real.
Usually, the Gregorian exit pupil is situated not far from the primary focus.
Such a position allows us to place a correcting optical element directly in the
exit pupil, providing efficient correction of aberrations of a two-mirror
system without auxiliary optics.

At first glance, the superposition of wide light beams near the primary focus
prevents to imposing a lens corrector in the Gregory system (see Fig.~1).
However, as shown below, it is possible to avoid additional obscuration, if we
make a hole at the center of the front lens of the corrector and shift the rear
its part to the primary mirror. As a result, we obtain a wide-field
catadioptric system that combines compactness with simple shape of the optical
surfaces\footnote{Strictly speaking, since both mirrors are optimized along
with the lens corrector, we deal not with a corrector to the pre-designed
aplanatic Gregory telescope, but with a new catadioptric system.}. The system
provides the sub-arcsecond field of view ${\sim2^\circ.5}$ in diameter even at
all-spherical corrector. It is worth noting that the spherical corrector for
the Gregory system repeats the lens corrector that was proposed earlier for a
single hyperbolic mirror (Terebizh,~2004). Subsequent aspherization of some
lens surfaces allows to achieve the field about $3^\circ$ in diameter.

In present paper, we discuss the Gregorian corrector with an example of a 6.5-m
telescope with the focal ratio $\phi \equiv F/D \simeq 1.9$. The effective
focal length of the telescope, $F \simeq 12.4$~m, allows to fit resolutions of
the optics and actual light detectors. The corrector consists of five lenses
made of the same, virtually any material. At use fused silica and simple
coating as a $MgF_2$ single layer, the telescope light transmission reaches
70\%. The detector window has some optical power, the field is slightly curved,
that is quite allowable, taking into account its linear size: approximately
0.5~m.

\begin{figure}[t]   
   \centering
   \includegraphics[width=\textwidth]{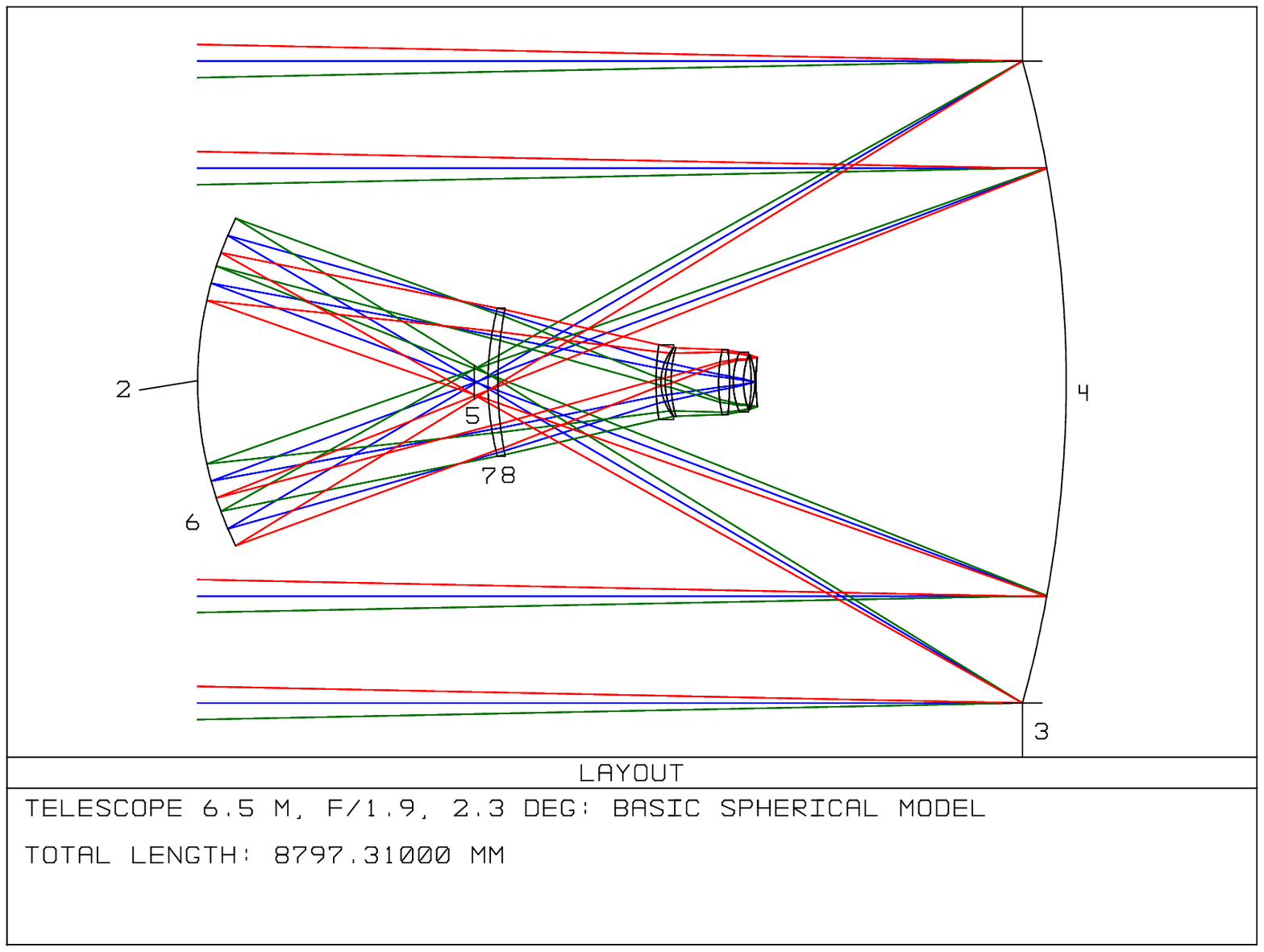}
   \caption*{Optical layout of the 6.5-m telescope with the spherical basic
   corrector. Ordering of the surfaces corresponds to the Table~2.}
\end{figure}

\section*{Basic system of the 6.5-m  telescope}

In a basic configuration of the telescope (Figs.~1,~2), the primary and
secondary mirrors are pure ellipsoids, while the lens surfaces are spherical.
General performance of the telescope for a case, when the field of view is
$2^\circ.3$ in diameter, is given in the Table~1; the parameters of its optical
layout are specified in the Table~2. At the description of the optical layout
we have introduced, for convenience, a fictitious surface No.~5, located close
to the paraxial primary focus.

Figures~3 and~4 show image quality provided by the basic system, optimized for
the field of view $2^\circ.3$. The circle diameter that contains 80\% of a
stellar image energy (designed, as usually, by $D_{80}$), varies within the
waveband $0.35-0.90\,\mu$m from $\sim0''.25$ at the center of the field up to
$\sim0''.50$ at its edge. The linear coefficient of central obscuration $\eta =
0.51$, so the effective aperture diameter of the system is 5.6~m. The fraction
of vignetted rays enlarges from the center of field to its edge less than on
2\%.

\begin{center}    
 \small{
\begin{tabular}{|l|c|c|c|}
\multicolumn{4}{l}{\textit{Table~1. Basic system of the 6.5-m telescope}}\\
\hline
    & \multicolumn{3}{|c|}{}\\
    & \multicolumn{3}{|c|}{Waveband, $\mu$m} \\
 \multicolumn{1}{|c|}{Parameter} & \multicolumn{3}{|c|}{}\\
 \cline{2-4}
    &&&\\
    & 0.35 -- 0.45 & 0.54 -- 0.66 & 0.70 -- 0.90 \\
    &&&\\
\hline
 Field of view  & \multicolumn{3}{|c|}{}\\
 \qquad Angular  & \multicolumn{3}{|c|}{$2^\circ.3$} \\
 \qquad Linear   & \multicolumn{3}{|c|}{498 mm} \\
\hline
 Effective focal length, mm &12370.7 &12368.9 &12367.8 \\
\hline
 Relative focal length & \multicolumn{3}{|c|}{1.903} \\
\hline
 Length of the system  & \multicolumn{3}{|c|}{8797.3 mm} \\
\hline
 Scale, $\mu$m/arcsec  &59.97 &59.97 &59.96 \\
\hline
 Relative vignetting    & \multicolumn{3}{|c|}{} \\
 at the edge of field   & \multicolumn{3}{|c|}{1.7\,\%} \\
\hline
 Variation of the RMS
   &$5.9-14.7\,\mu$m &$5.0-8.0\,\mu$m &$5.4-11.0\,\mu$m \\
 image radius over field
   &$0''.10-0''.25$ &$0''.08-0''.13$ &$0''.09-0''.18$ \\
\hline
 Variation of $D_{80}$ from the
   &$15.8-31.6\,\mu$m &$14.4-22.1\,\mu$m &$16.8-29.7\,\mu$m\\
 center to the edge of field
   &$0''.26-0''.52$ &$0''.24-0''.37$ &$0''.28-0''.50$\\
\hline
 Transmission with the           &&&\\
 single layer of $MgF_2$         &$0.70$ &$0.71$ &$0.70$\\
\hline
 Maximum distortion  & 0.27\,\% & 0.28\,\% & 0.29\,\%\\
\hline
 The lens surfaces & \multicolumn{3}{|c|}{All spheres} \\
\hline
\end{tabular}
 }
\end{center}

\bigskip

The relative focal length of the primary mirror, $\phi_1 \equiv F_1/D_1$, is
about 0.92, for the secondary mirror we have $\phi_2 = 0.56$. Let us remind,
for comparison, relative focal lengths of three mirrors of the Large Synoptic
Survey Telescope (LSST) according to Seppala~(2002): $1.057$, $0.914$ and
$0.774$; the light diameters are $8.40$~m, $3.37$~m and $5.44$~m, respectively.
These mirrors are aspherics of the 6th--10th orders, the secondary mirror is
convex. The primary mirrors of the Large Binocular Telescope (LBT) of diameter
8.4~m have $\phi_1 = 1.14$ (Hill, 1996; Salinari, 1996).

\begin{figure}[t]   
   \centering
   \includegraphics[width=0.90\textwidth]{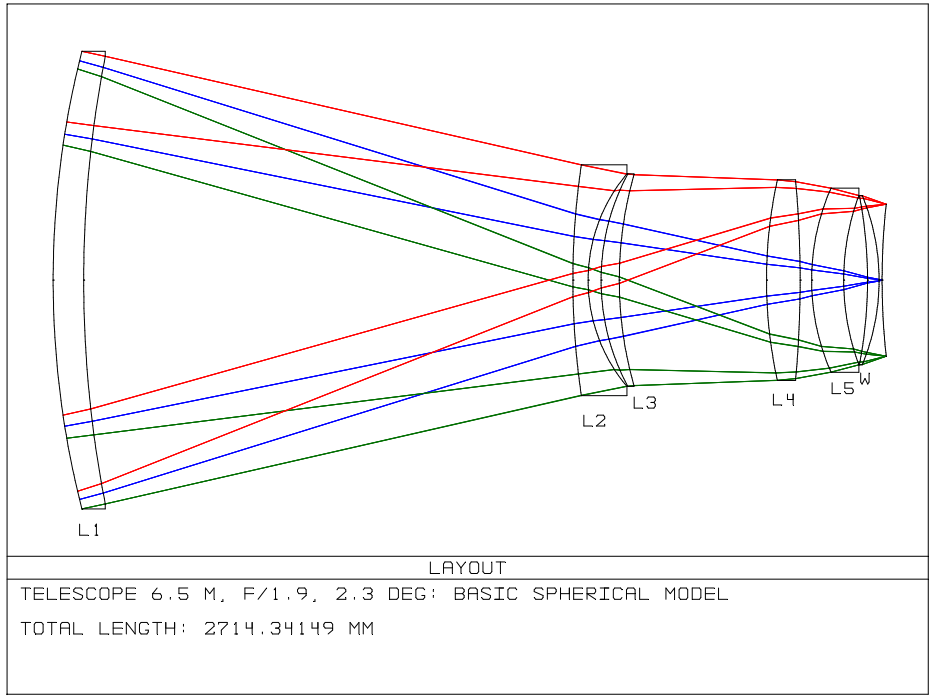}
   \caption*{Optical layout of the basic corrector. Numbers of lenses\\
    correspond to the Table~2, the detector window is marked by `W'.}
\end{figure}

\bigskip

According to the above data, we expect no specific problems at manufacturing of
monolithic primary for the proposed here system, but the secondary mirror seems
to be dangerously fast. However, we have to take into account that both mirrors
are the concave pure ellipsoids, which can be controlled during manufacturing
with the aid of the well-known and reliable methods. The other aspect of the
problem under discussion concerns the customary nowadays practice of including
the secondary mirrors to the active optics systems (e.g., on the LBT). These
mirrors properly change their shape under action of actuators, and have very
complicated form at each moment of time. For this reason, the initial form of a
secondary mirror in active system is not obliged to follow the exact design.
Further, it is known that making of fast mirrors becomes strongly simpler at
use of the mosaic technology (see, e.g., Mountain and Gillett,~1998;
Wilson,~1999). Let us notice, at last, that the f/number of a secondary depends
upon a set of general characteristics of a telescope, and, in case of need, one
can initially choose the characteristics in such a way that $\phi_2$ gets the
desirable range. All said above allows to hope, that manufacturing of a
secondary mirror for the proposed system is within modern technological
abilities.

\begin{center}   
 \small{
\begin{tabular}{|c|c|c|c|c|c|c|}
 \multicolumn{7}{l}{\textit{Table 2. Parameters of the basic 6.5-m
  telescope, optimized for the field $2^\circ.3$}}\\
\hline
 Number  &         &Curvature &Thickness &      &Light    &Conic \\
 of the  &Comments &radius    &(mm)      &Glass &diameter &constant \\
 surface &         &(mm)      &          &      &(mm)     & \\
\hline
 1  &Screen &$\infty$   &0 &--- &3315.00 & 0\\
 2  &Vertex of &&&&&\\
    & secondary &$\infty$ &8355.89 &---  & 0  &0\\
 3  &Aperture &&&&&\\
    &diaphragm  &$\infty$   &441.42     &---  &6500.00  & 0\\
 4  &Primary &$-11993.55$ &$-5995.78$ &Mirror &6500.00 &$-0.866870$\\
 5  &Primary focus &$\infty$ &$-2801.53$ &--- &338.43 &0\\
 6  &Secondary &$3728.74$ &$2941.53$ &Mirror &3314.54 &$-0.194002$\\
 7$^{1)}$   &L1       &$3034.14$   &$100.00$  &FS$^{3)}$
       &1500.00 & 0\\
 8$^{2)}$  &         &$3835.23$   &$1601.82$ &---   &1466.74 & 0\\
 9    &L2       &$2587.74$   &$50.00$   &FS &758.08 & 0\\
 10   &         &$541.32$    &$42.41$   &---   &697.64 & 0\\
 11   &L3       &$736.49$    &$59.65$   &FS    &698.08 & 0\\
 12   &         &$1268.16$   &$481.986^{4)}$ &---  &693.44 & 0\\
 13   &L4       &$1532.40$   &$110.00$  &FS    &655.58 & 0\\
 14   &         &$-3638.17$  &$37.74$   &---   &644.04 & 0\\
 15   &L5       &$747.81$    &$105.00$  &FS    &601.65 & 0\\
 16   &         &$800.48$    &$78.00$   &---   &553.91 & 0\\
 17   &Window   &$-1452.45$  &$37.71$   &FS    &553.05 & 0\\
 18   &         &$-727.49$   &$10.00$   &---   &551.45 & 0\\
 19   &Detector &$2322.90$   &          &      &498.01 & 0\\
\hline
\end{tabular}
 }
\end{center}
 $^{1)}$  {\small The hole of 427.7~mm in diameter}.\\
 $^{2)}$ {\small The hole of 542.3~mm in diameter}.\\
 $^{3)}$ {\small $FS$~-- fused silica}.\\
 $^{4)}$ {\small The visual waveband is meant. The distances
 for the blue and red wavebands are 482.011~mm and 481.936~mm,
 respectively}.

\bigskip

\begin{figure}[t]   
   \centering
   \includegraphics[width=0.75\textwidth]{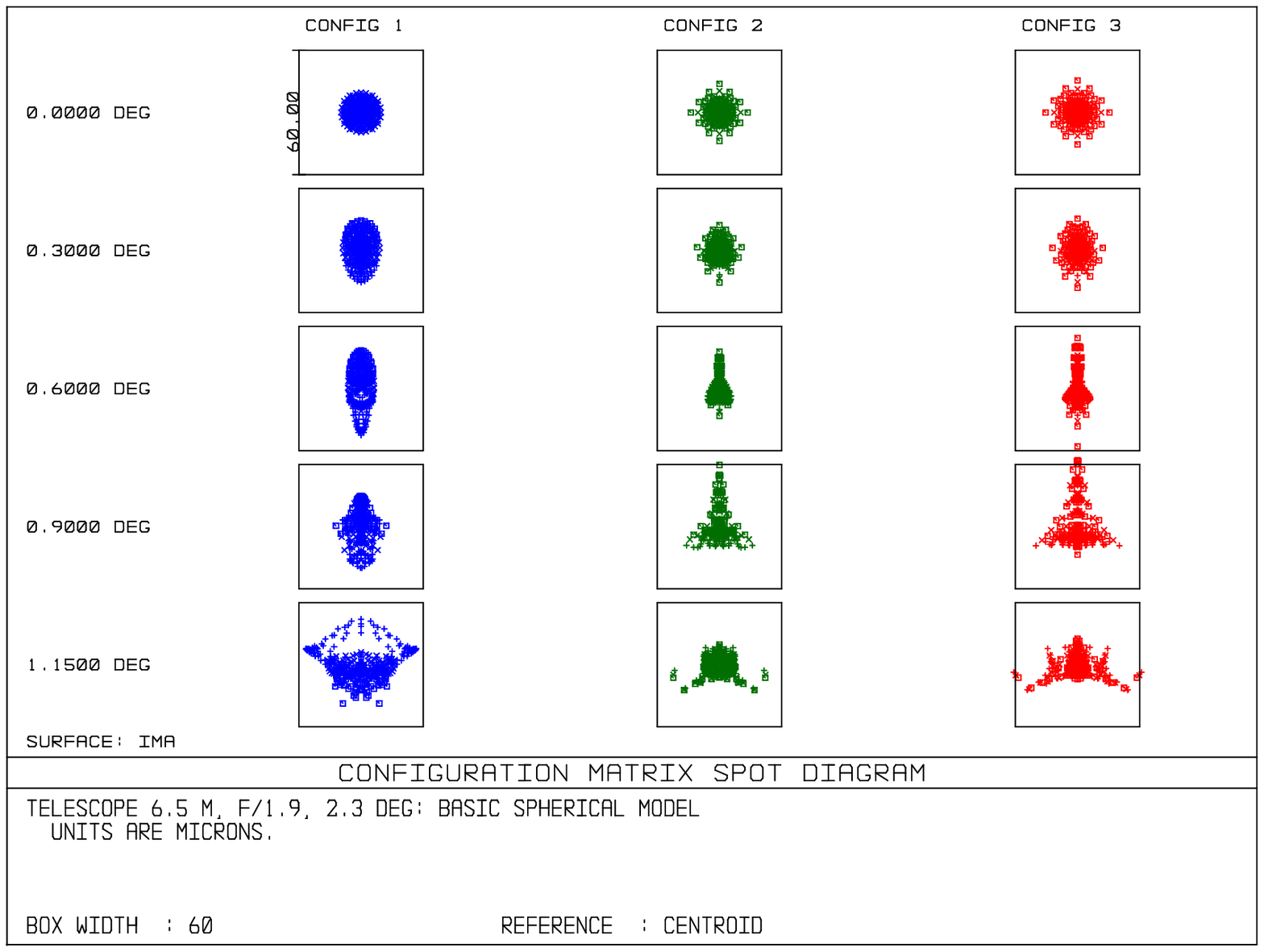}
   \caption*{Spot diagrams of the 6.5-m telescope with spherical
    basic corrector\\ for the field angles $0;\, 0^\circ.3;\, 0^\circ.6;\,
    0^\circ.9$ and $1^\circ.15$ (the rows). The columns correspond to
    the wavebands $0.35-0.45\,\mu$m, $0.54-0.66\,\mu$m and
    $0.70-0.90\,\mu$m, respectively. The box width is $1''$ ($60\,\mu$m).}
\end{figure}

\begin{figure}[t]   
   \centering
   \includegraphics[width=0.75\textwidth]{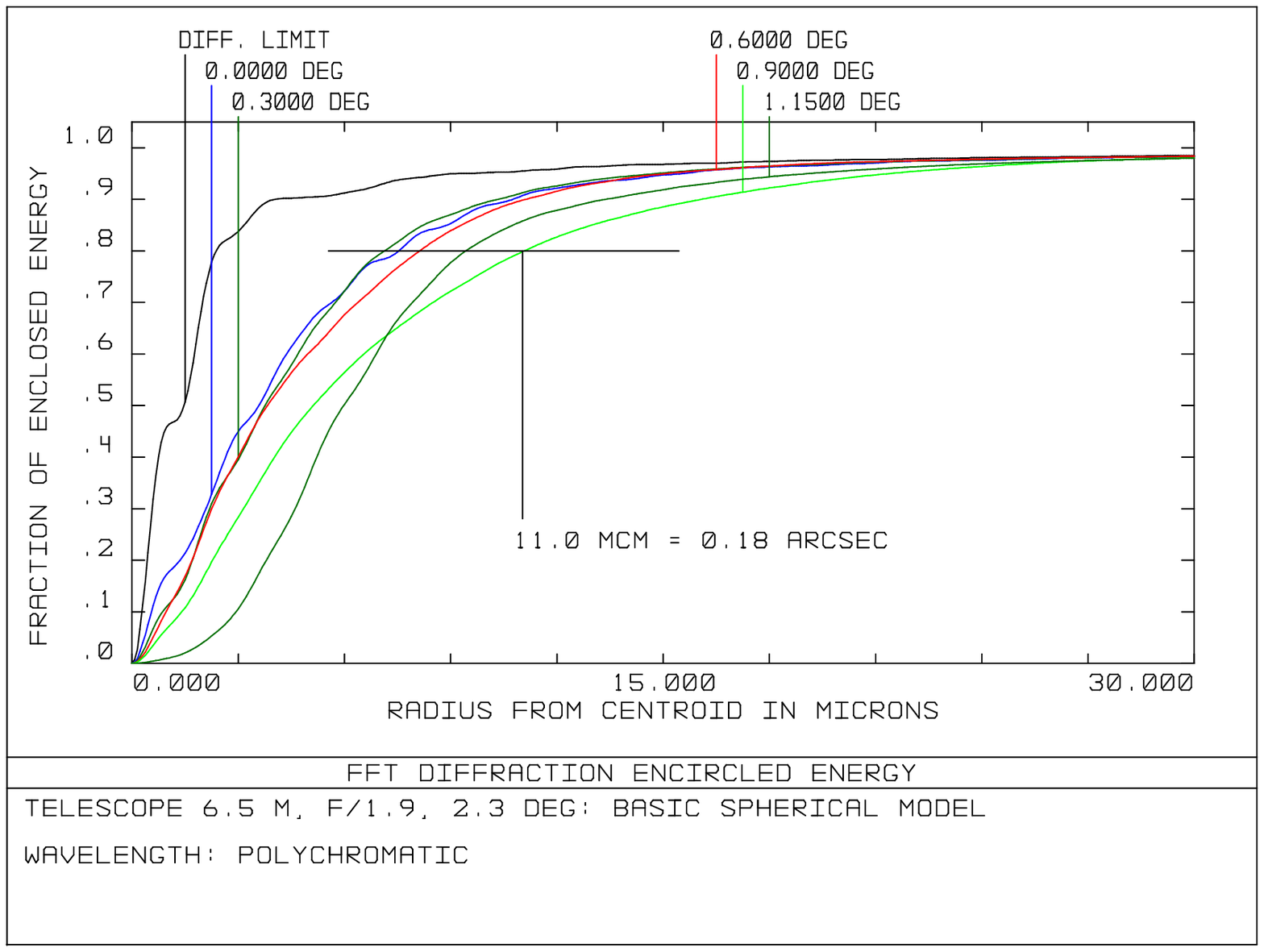}
   \caption*{Basic telescope: Integral energy distribution along radius
   in a star image for the waveband $0.54-0.66\,\mu$m and the field
   angles $0;\, 0^\circ.3;\, 0^\circ.6;\, 0^\circ.9$ and $1^\circ.15$.
   The similar distribution in the diffraction-limited image and the
   $80\%$-level are also shown.}
\end{figure}

\begin{figure}[t]   
   \centering
   \includegraphics[width=0.75\textwidth]{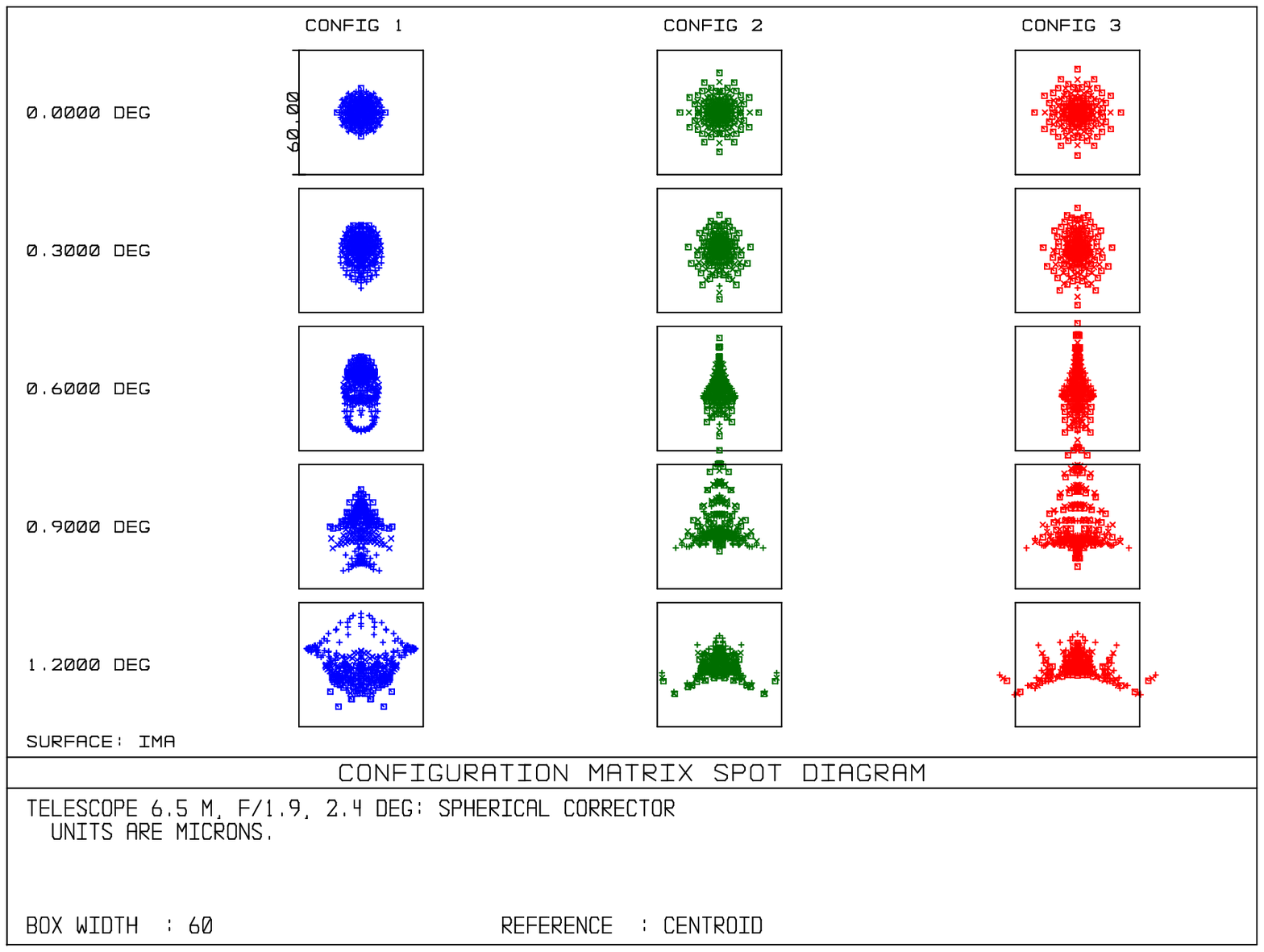}
   \caption*{Spot diagrams of the 6.5-m telescope with spherical
    corrector for the\\ field angles $0;\, 0^\circ.3;\, 0^\circ.6;\,
    0^\circ.9$ and $1^\circ.2$ (the rows). The columns correspond to the wavebands
    $0.35-0.45\,\mu$m, $0.54-0.66\,\mu$m and $0.70-0.90\,\mu$m, respectively.
    The box width is $1''$ ($60\,\mu$m).}
\end{figure}

\begin{figure}[t]   
   \centering
   \includegraphics[width=0.75\textwidth]{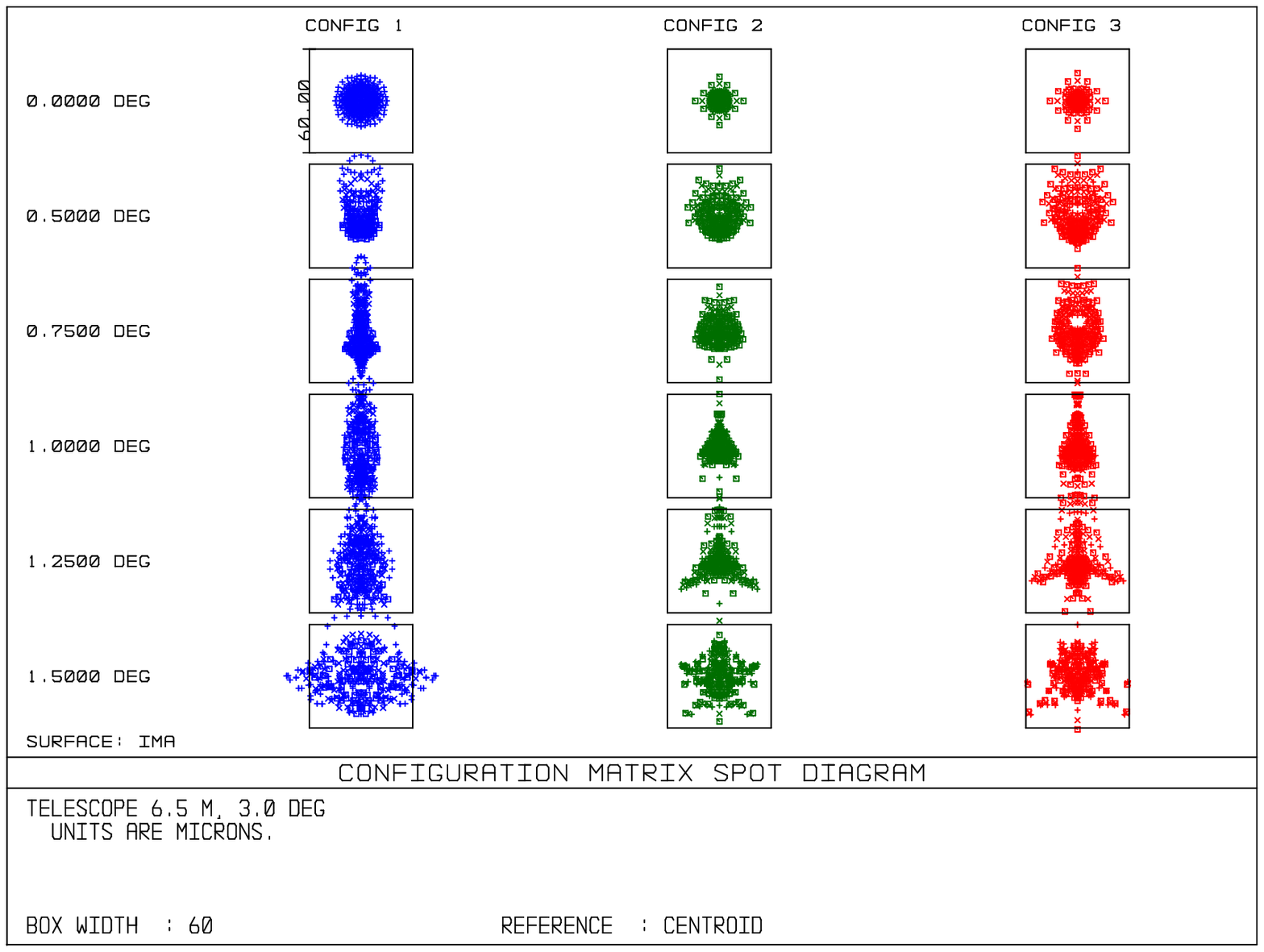}
   \caption*{Spot diagrams of the 6.5-m telescope with aspheric
    corrector for the\\ field angles $0;\, 0^\circ.5;\, 0^\circ.75;\,
    1^\circ.0$, $1^\circ.25$ and $1^\circ.50$. The columns correspond to the
    wavebands $0.35-0.45\,\mu$m, $0.54-0.66\,\mu$m and $0.70-0.90\,\mu$m,
    respectively.
    The box width is $1''$ ($60\,\mu$m).}
\end{figure}

\bigskip

The aperture diameter and general characteristics of the telescope were mainly
determined by condition, that the diameter of the front corrector lens L1 (see
Fig.~2) is no more than 1.5~m (the front lens of the LSST corrector is 1.34~m).
A central cone-shaped hole should be made in L1 for passage of light beam
reflected by the primary mirror. It is possible to manage without the hole,
supposing double passage of light through the lens L1, but the image quality in
a correspondingly optimized system is not so high.

Note that lens sizes are close to those in the prime focus corrector to a
single hyperbolic mirror of 4~m in diameter (Terebizh,~2004). Thus, application
of a Gregory corrector allows to essentially increase the telescope aperture~--
in this case from 4.0~m up to 6.5~m,~-- while the system length has decreased
from 10.8~m down to 8.8~m.

The focal surface of about 0.5~m in diameter is a convex sphere of curvature
radius about 2.3~m. The corresponding sag at the field edge is 13.4~mm.
Relatively small field curvature does not prevent to placing a set of matrix
detectors, which own sizes are less than $\sim25$~mm.

There are a few ways to control focusing at change of the spectral range; we
choose, as an illustration, variation of the distance between the third and
forth lenses. Namely, according to the Table~2, one should shift the rear part
of the corrector only by $+25\,\mu$m and $-50\,\mu$m to turn from the visible
range to the blue and red wavebands, respectively.

Under the pixels size ${\sim 15\,\mu}$m, which is typical for the modern CCD's,
one pixel corresponds to $0''.25$. Thus, about $1-2$ pixels cover a star image
of $D_{80}$ in diameter, and we may consider as feasible the matching of
resolution of the optical system with that of the detector and the atmosphere
image quality.

The telescope light transmission has been estimated, assuming the simple
coatings~-- the single  $\lambda/4$ layer of $MgF_2$. Of course, the modern
multi-layer coatings will ensure best transmission of light.

It is important to note, that the corrector is close to an afocal system, so
the optical power of the telescope is determined mainly by its mirrors.
Evidently, just that feature allows to avoid chromatism and, as a consequence,
to attain the large field of view. This general principle is true also for
other catadioptric systems.

\section*{Extending the field of view}

The all-spherical corrector provides the field up to about $2^\circ.5$. Fig.~5
depicts, as an example, the spot diagrams for the field $2^\circ.4$ in
diameter. Comparison with the Fig.~3 shows that the image quality has worsened
only a little. Further extension of the field meets difficulties caused, first
of all, by accepted here restriction of the sizes of the corrector front lens.

As is well known, there is a quite simple, from a designer's point of view, way
to attain the more wide field of an optical system: the aspherization of the
all or some surfaces. Certainly, this way complicates technical realization of
the system. In particular, the tolerances becomes much more hard, so both the
fabrication and use of the telescope is laborious. Ultimately, these factors
have an essential effect on the cost of the telescope. Nevertheless, many large
telescopes that are now in progress include polynomial aspherics up to 10th
order. In our case, aspherization of some surfaces of the lens corrector,
namely, adding the terms of 4th, 6th and 8th orders, provides the sub-arcsecond
field of view about $3^\circ$ in diameter (Fig.~6).

Even the wider field of view is attainable by applying the polynomial aspherics
not only on the lenses, but also onto the (concave) mirrors of the system. We
shall not consider here this opportunity, as now the main task is to give the
general description of the lens corrector for a quasi-Gregory telescope.

\section*{Concluding remarks}

Let us estimate the \textit{throughput}\footnote{\textit{\'Etendue} (Fr.)} $E$
of the proposed telescope and, for comparison, that of the LSST and a 4-m
one-mirror telescope. The frequently used now parameter $E$ is defined as
product of the telescope effective area by the solid angle, corresponding to
its field of view. Table~3 gives approximate values of $E$ for the two field
sizes. All systems under consideration include the lens field correctors.
Obviously, to continue discussion it is necessary to take into account also a
number of concomitant factors, as that: a reality of manufacturing of the
optical surfaces of required form, the tolerances on temporal stability of the
whole set of parameters, the operation cost of a telescope etc.

\begin{center}
 \small{
\begin{tabular}{|l|c|c|}
\multicolumn{3}{l}{\textit{Table 3. Throughput of some telescopes
 $(m^2\,deg^2)$}}\\
\hline
  & \multicolumn{2}{|c|}{}\\
  \multicolumn{1}{|c|}{} & \multicolumn{2}{|c|}{Field of view} \\
  \multicolumn{1}{|c|}{Telescope} & \multicolumn{2}{|c|}{}\\
  \cline{2-3}
  & $2^\circ.3$ & $3^\circ.0$ \\
\hline
One-mirror 4.0-m telescope        &&\\
with a prime-focus corrector      & 46    & 78  \\
Two-mirror 6.5-m Gregory with     &&\\
the corrector at the exit pupil   & 102   & 170 \\
Three-mirror 8.4-m LSST with      &&\\
the three-lens corrector          &  --   & 264 \\
\hline
\end{tabular}
 }
\end{center}

\bigskip

It is worth mentioning, that the two-mirror telescope alone, taken as a part of
the considered here catadioptric system, provides the image of an axial
point-like object of $0''.15$ in diameter, but a quarter of degree off-axis
image is already $6''$ in diameter. Nearly the same characteristics has the
aplanatic version of a Gregory telescope without lens corrector.

In the Introduction, we have touched on the attractive features of the Gregory
telescope: reality of its exit pupil and concave form of the secondary mirror.
Let us remind also, that it is much easier in the Gregory system to design an
efficient baffles than in the Cassegrain system (Terebizh,~2001).

Naturally, the described above 6.5-m telescope is only one of examples; our
main purpose was to attract attention to possibility of versatile use of the
central and periphery zones of the front lens of the corrector placed in the
exit pupil of a two-mirror Gregory telescope. Proceeding from the basic
configuration, it is possible to design systems with account of the particular
conditions and auxiliary optics (e.g., filters and the atmospheric dispersion
corrector). The scaling of the system to smaller diameters does not meet
problems, but scaling to larger diameters causes increasing of the corrector
front lens.

\bigskip

The author is grateful to V.V.~Biryukov for useful discussions.

\end{document}